\documentclass[letterpaper]{aa}
\usepackage{graphics}
\usepackage{epsfig}
\begin{document}
\thesaurus{08.12.3, 08.06.02, 09.08.1, 09.13.2, 13.09.3, 13.19.3}

\title{The luminosity function of galactic ultra-compact H\,{\sc ii}
regions and the IMF for massive stars \thanks{based partly on results
collected at the European Southern Observatory, La Silla, Chile}}

\titlerunning{The luminosity function of galactic UCH\,{\sc ii} regions}

\author
{S.~Casassus\inst{1,2} \and L.~Bronfman\inst{1} \and J.~May\inst{1} \and L.--\AA.~Nyman\inst{3,4}}
\offprints{L.~Bronfman}
\institute{Departamento de Astronom\'{\i}a, Universidad de Chile, Casilla 36-D, Santiago, Chile 
\and Astrophysics,  University of Oxford, Keble Road, Oxford OX1 3RH, UK 
\and SEST, ESO-La Silla, Casilla 19001, Santiago 19, Chile
\and Onsala Space Observatory, S--439 92 Sweden}
\date{received;      accepted}
\maketitle
\begin{abstract}

The population of newly formed massive stars, while still embedded in
their parent molecular clouds, is studied on the galactic disk scale.
We analyse the luminosity function of IRAS point-like sources,
with far-infrared (FIR) colours of ultra-compact H\,{\sc ii} regions,
that have been detected in the CS(2--1) line - a tracer of high
density molecular gas.  The FIR luminosities of 555 massive star
forming regions (MSFRs), 413 of which lie within the solar circle, are
inferred from their fluxes in the four IRAS bands and from their
kinematic distances, derived using the CS(2--1) velocity profiles.
The luminosity function (LF) for the UCH\,{\sc ii} region candidates
shows a peak well above the completeness limit, and is different
within and outside the solar circle (96\% confidence level). While
within the solar circle the LF has a maximum for
$2\,10^{5}\,L_{\odot}$, outside the solar circle the maximum is at
$5\,10^{4}\,L_{\odot}$. We model the LF using three free parameters:
$-\alpha$, the exponent for the initial mass function (IMF) expressed
in $\log(M/M_{\odot})$; $-\beta$, the exponent for a power law
distribution in $N^{\star}$, the number of stars per MSFR; and
$N^{\star}_\mathrm{max}$, an upper limit for $N^{\star}$.  While
$\alpha$ has a value of $\sim 2.0$ throughout the Galaxy, $\beta$
changes from $\sim 0.5$ inside the solar circle to $\sim 0.7$ outside,
with a maximum for the number of stars per MSFR of $\sim$650 and
$\sim$450 (with $1 \leq M/M_{\odot} \leq 120$).  While the IMF appears
not to vary, the average number of stars per MSFR within the solar
circle is higher than for the outer Galaxy.
  
\end{abstract}

\section{Introduction}

A new tool is now available to probe the population of recently formed
massive stars, while still embedded in their parent clouds. Such stars
are surrounded by a compact H\,{\sc ii} region, with an ionization
front working outwards into the cloud. Regions undergoing massive star
formation contain one or more ultra-compact H\,{\sc ii} (UCH\,{\sc
ii}) regions, and possibly more evolved H\,{\sc ii} regions.  Bronfman
et al. (1996, BNM) completed a survey in CS(2--1) towards IRAS point
sources satisfying the Wood \& Churchwell (1989a) far-infrared (FIR) colour
criteria for UCH\,{\sc ii} regions. The CS molecule is a tracer of
high density molecular gas; a CS(2--1) detection strengthens the
UCH\,{\sc ii} region identification and provides kinematic
information. BNM detected 843 sources (hereafter IRAS/CS sources),
whose azimuthally averaged galactic distribution is presented in
Bronfman et al. (2000, BCMN). In this work we construct the luminosity
function (LF) for the IRAS/CS sources, and show that it presents
significant variations with galactocentric radius. We interpret the
shape of the IRAS/CS sources LF and its variations in terms of an
ensemble of massive star forming regions. The stellar content of
IRAS/CS sources is characterised by the number of stars and their mass
spectrum which, because of the youth of the systems, is taken to
represent closely their initial mass function (IMF).

The use of young tracers to derive the IMF minimises the dependence on
modelling, which afflicts most previous studies. The standard approach
to determining the high mass end of the IMF has been through O and B
star counts in the solar neighbourhood (e.g. Miller \& Scalo 1979,
Lequeux 1979). The average mass spectrum of newly formed stars is
taken as a power law, and the value of the exponent characterises the
IMF. These approaches require assumptions on the star formation
history, and have the drawback of not probing the whole galactic
disk. Garmany et al. (1982) addressed the question of large scale
variations in the IMF exponent, and they favour a decrease with
galactocentric radius (although their result has been reinterpreted,
e.g. Massey 1998). A different tool was used by V\'{a}zquez \&
Feinstein (1989), who linked the variations in the open cluster
luminosity function (Burki 1977) to variations in the IMF index.

Young objects such as H\,{\sc ii} regions have also been used to study massive
stars in the galactic context.  McKee \& Williams (1997, MW97)
characterised the population of newly formed massive stars through the
luminosity function of OB associations. The ionising luminosity
absorbed by the gas surrounding OB associations can be traced through
the radio flux of H\,{\sc ii} regions, and the frequency distribution in
luminosity of H\,{\sc ii} regions is in turn a function of the number of
exciting stars and their masses.  However, MW97 a-priori fixed the
shape of the IMF and the distribution for the number of exciting
stars. This method is also strongly model dependent due to the lack of
complete information on galactic H\,{\sc ii} regions, and the fact that the
population of H\,{\sc ii} regions is not homogeneous in age.  Comer\'{o}n \&
Torra (1996, CT96) also analysed the galactic disk distribution of newly
formed massive stars, through a sample of IRAS point sources with
colours of UCH\,{\sc ii} regions. But they did not use kinematic information
to derive the galactic distribution of the UCH\,{\sc ii} regions, and their
work is based on a position-independent LF.

The aims of this work are to present the FIR LF of massive star
forming regions still embedded in their parent clouds; to investigate
the LF large scale variations; and to analyse the LF in terms of the
embedded stars' mass spectrum. In Sect. \ref{sec:obs} we construct
the LF of IRAS/CS sources for different sectors in the galactic disk,
and show that it presents significant variations with galactocentric
radius.  In Sect. \ref{sec:model} we analyse the stellar population
underlying the IRAS/CS sample through a simple model based on a
synthetic ensemble of massive star forming regions (MSFRs). A search
in parameter space is conducted in Sect. \ref{sec:paramspace}. The
results of this analysis and the questions of the number and the mass
spectrum of stars per MSFR will be addressed in Sect.
\ref{sec:results}. In particular, we will show that the mass spectrum
of newly formed massive stars is constant across the galactic disk,
although the average number of stars per MSFR is lower outside the
solar circle. Section \ref{sec:lifetimes} is a brief estimate of the
fraction of lifetime O stars spend in the embedded phase. In Sect.
\ref{sec:conclusion} we summarise our conclusions.

\section{FIR luminosity function of UCH\,{\sc ii} regions}
\label{sec:obs}
The velocity information provided by the CS(2--1) observations from BNM
allows, through the adoption of a rotation curve, to derive their
galactocentric distances, and hence their heliocentric distances and
luminosities. The galactic disk is assumed to be in circular motion,
with $R_{\circ}=8.5\,$kpc and $V_{\odot}=220$ km\,s$^{-1}$. For the region 
of the Galaxy outside the solar circle (outer Galaxy) the procedure is a
simple coordinate transformation from galactic longitude, latitude and
velocity ($l,\, b,\, V_{\rm lsr}$) to galactocentric radius, height
over the plane and azimuth ($R,\, z,\, \theta$). But such a
transformation is bivalued for the region within the solar circle (inner Galaxy). There is a heliocentric distance ambiguity such that, unless a 
source lies just on the subcentral point (the tangent point to a 
galactocentric ring for a given longitude), there are two points along 
the line of sight, at the same distance on both sides of the subcentral
point, that have the same line of sight velocity.


The method we used to resolve the distance ambiguity is described in
BCMN. It is a statistical method that consists in weighting the near
and far distances with a normal distribution in height over the
galactic plane.  Each source is assigned an effective luminosity,
which is the weighted average of the near and far luminosities. The
centroid and width of the vertical distribution, $Z_{\circ}(R)$ and
$Z_{\frac{1}{2}}(R)$, are determined through an iterative process for
galactocentric bins $0.1\,R_{\circ}$ wide.  A consistency check for
this method can be found below in this Section.

The kinematic distances are not reliable in the direction of the
galactic centre, and also when the line of sight velocities are of the
same order as the non-circular velocity components. Therefore, we
excluded from the present analysis all sources within $\pm 10^{\circ}$
of the galactic centre, and within $\pm 5^{\circ}$ of the galactic
anti-centre, as well as sources with $|V_{\rm lsr}|\leq 10\, {\rm
km}\, {\rm s}^{-1}$. The resulting range in galactocentric radius,
where the disk is properly sampled, excludes the solar circle. We
restricted the analysis to sources with $0.3~ \leq R/R_{\circ} \leq
0.9$ and $1.1  \leq R/R_{\circ} \leq 1.6$.

The sources close to the subcentral points are an important
consistency check of our method to resolve the distance ambiguity
within the solar circle. The kinematic distance of a source at the
subcentral point and in pure circular motion about the galactic
centre is uniquely determined.  The subcentral sources sample was
defined as the subset with line of sight velocities no more than 10
~km\,s$^{-1}$ different in absolute value from the terminal velocity
(the maximum velocity expected for a given longitude in the case of
circular rotation).

\begin{figure}
 \resizebox{8cm}{!}{\epsfig{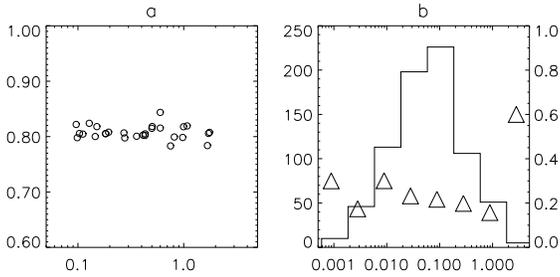}} \caption{a) The ratio
 of the fluxes published by WC89 to $F_{IRAS}$, as a function of
 $F_{IRAS}$ in 10$^{-9}$ W m$^{-2}$. b) The number of sources (left
 hand scale) as a function of $F_{IRAS}$ in 10$^{-9}$ W m$^{-2}$. The
 proportion of sources with an upper limit in the 100$\,\mu$m band is
 shown in triangles (right hand scale)}\label{fig:FIR}
\end{figure}

We estimate the far infrared flux of an IRAS/CS source by
summing over the four IRAS bands,
\begin{equation}
\label{eq:Firas}
F_{IRAS}=\sum_{j=1}^{4} \,\nu \,F_{\nu}(j),
\end{equation}
where $F_{\nu}(j)$ are the IRAS band flux densities, as listed in the
IRAS Point Source Catalog (1985).  In order to test this
approximation we compared with the total fluxes reported for 53
UCH\,{\sc ii} regions by Wood \& Churchell (1989b, WC89, their Tables
17 and 18). Figure \ref{fig:FIR}a shows the ratio of the $F_{IRAS}$
fluxes (obtained by Eq. \ref{eq:Firas}) to the total fluxes of WC89
(which are integrated up to 100$\,\mu$m), as a function of $F_{IRAS}$.
Equation \ref{eq:Firas} overestimates the WC89 fluxes by about a
constant 20\%, but WC89 did not include a correction for the
100$\,\mu$m -- 1 mm flux, which they estimate could be as high as
50\%.  We thus expect Eq.  \ref{eq:Firas} to be a good estimate of the
total luminosity of IRAS/CS sources within 30\% (50\% -- 20\%).

Since we will average the luminosity function of UCH\,{\sc ii} regions
over large areas of the galactic disk, it is important to estimate the
minimum luminosity above which the disk is properly sampled. The
lowest flux $F_{IRAS}$ in the sample, $F_{IRAS}=6\, 10^{-13}$
W\,m$^{-2}$, corresponds to $\log(L_\mathrm{min}/L_{\odot})=3.6$ at a
distance of 15~kpc. Figure \ref{fig:FIR}b shows a histogram of the
total number of sources in our sample as a function of $F_{IRAS}$
(without the velocity filter $|V_{\rm lsr}|\leq 10\, {\rm km}\, {\rm
s}^{-1}$). The triangles show the fraction of sources with only upper
limits in the 100$\,\mu$m band (right hand scale). Could a significant
number of sources be missed by IRAS near the minimum detected
flux? A lower $F_{IRAS}$ sensitivity would be hinted at by an
increased fraction of upper limits in the reported IRAS fluxes,
which is not the case. However, the higher far-IR background towards
the central regions of the Galaxy results in a completeness limit of
$\log(L_\mathrm{min}/L_{\odot})=4.5$ at 8.5~kpc, within $-10<l<10$,
$-0.3<b<0.3$. But over a broader longitude range,
$\log(L_\mathrm{min}/L_{\odot})=4.0$ at 8.5~kpc, within $-60<l<60$,
$-0.3<b<0.3$. As the luminosity functions of the subcentral sources
(which are all within 8.5~kpc of the Sun) is in good agreement with
that of the whole inner Galaxy (see below), we take
$\log(L_\mathrm{min}/L_{\odot})=4$ as the completeness limit of the
IRAS/CS sample\footnote{The CS(2--1) detection requirement has
been checked (BCMN) not to introduce additional biases within
8.5~kpc}.


\begin{figure}
 \resizebox{8cm}{!}{\epsfig{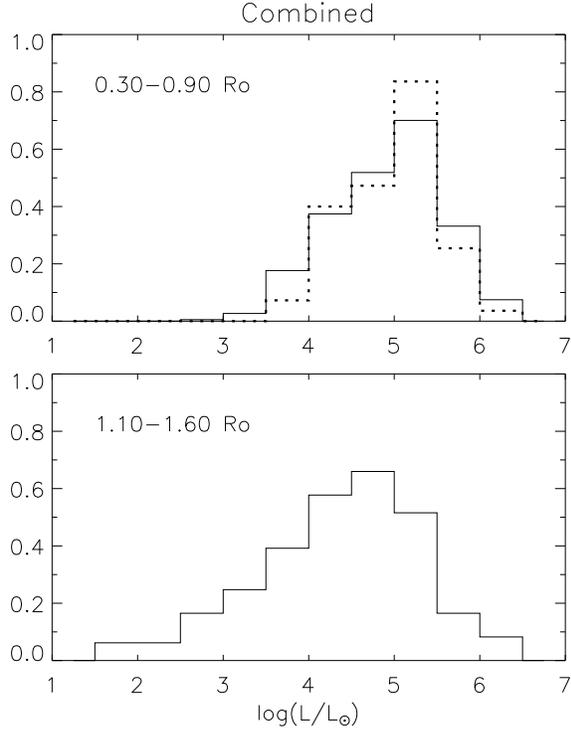}} \caption{ The
 luminosity function for galactic UCH\,{\sc ii} regions, from the IRAS/CS
 sample.  The whole disk was divided at the solar circle, the upper
 and lower plots correspond to the inner and outer Galaxy LFs,
 computed with 413 and 142 sources respectively. In the upper plot the
 inner Galaxy LF derived from the effective luminosities is shown in
 solid line, while the thick dotted line is the LF for sources near
 the subcentral points (57 sources). The distributions in these plots
 are normalised so that the areas under the histograms is one over
 $\log(L/L_{\odot}$)$>$4. Shot noise gives 1-$\sigma$ error bars on
 the subcentral LF of $\sim$25\%}\label{fig:LF_gal}
\end{figure}

The LF of the whole sample of IRAS/CS sources, our main observational
result, appears to be significantly different inside and outside the
solar circle; in Fig. \ref{fig:LF_gal} we distinguish between
$R<R_{\circ}$ and $R>R_{\circ}$.  The LFs cover a very wide range in
luminosity, over three orders of magnitude, which allows using
logarithmic luminosity bins corresponding to a factor of 300\%.
Within the solar circle the LF obtained using the effective
luminosities is confirmed to be a good estimate of the actual LF
through its close similarity with the LF of the sources near the
subcentral points, shown in dotted line\footnote{A total of 57 sources
were used to calculate this LF. Sources with galactocentric radii
larger than $0.8\,R_{\circ}$ were excluded: our definition of the
subcentral source sample would otherwise include most sources in that
region, because the non-circular velocity components dominate the
spread about the terminal velocity}. For comparison, placing all the
sources at the `near' or `far' kinematic distance changes the peak of
the LF as a function of logarithmic luminosity from 4.25 to 5.75,
while the LF obtained using the effective luminosities peaks at
5.25. The good match with the subcentral source sample LF lends
strength to a comparison of the luminosity functions between the outer
and inner Galaxy, based on the effective luminosities.  We will refer
to the luminosity functions for the whole inner and outer Galaxy by
LF$^\mathrm{in}$ and LF$^\mathrm{out}$.

The dominant source of uncertainty in the LF is shot noise.  The
errors in the galactic disk surface FIR luminosity amount to about
10\% upwards, 20\% downwards (Fig. 3 in BCMN). The fractional error on
the FIR surface luminosity represents the typical fractional error on
the luminosity of one source. These errors stem from the IRAS
100$\,\mu$m band flux uncertainty, coupled with the kinematic distance
uncertainty due to non-circular motions of about 5~km\,s$^{-1}$.
Adding in quadrature the 30\% uncertainty related to the use of
Eq. \ref{eq:Firas}, we have an average error on the luminosity of a
source of at most 36\%.  Compared to the 300\% width of the luminosity
bins, a 36\% uncertainty is negligible, apart from a slight smoothing
effect without practical consequence.

The differences in the LFs inside and outside the solar circle are
statistically significant.  As a statistic for the difference between
the inner and outer LFs, we used a $\chi^{2}$ test which has the
following expression in this context,
\begin{equation}
\chi^{2} = \sum_{i} \frac{
({\rm LF}^\mathrm{in}_{i}-{\rm LF}^\mathrm{out}_i)^{2}}{\frac{{\rm
LF}^\mathrm{in}_{i}}{(N_\mathrm{in}{\Delta}\log L)^2}+\frac{{\rm LF}^\mathrm{out}_{i}}{(N_\mathrm{out}{\Delta}\log L)^2}},
\end{equation}
where we sum over the bins above the luminosity limit of
10$^{4}\,L_{\odot}$. The result is $\chi^{2} = 11.4$, or that
LF$^\mathrm{in}$ and LF$^\mathrm{out}$ are different at a significance
level of 95.6\% (with a $\chi^2$ distribution for 5 degrees of
freedom, corresponding to the number of bins with non-zero counts
above the luminosity limit).  For comparison, the same test applied to
the northern and southern\footnote{We refer to the sector with
longitudes less than $180 ^{\circ}$ as the northern Galaxy, while
longitudes greater than $180 ^{\circ}$ correspond to the southern
Galaxy} LFs inside the solar circle gives a probability of 45\% that
the distributions are different, so they are comparable relative to
the differences between the LFs inside and outside the solar
circle. Another application of this statistical test gives that
LF$^\mathrm{in}$ and the subcentral source LF are the same at a
confidence level of 87\%.

We emphasize the presence of a peak in the LF of IRAS/CS sources, well
above the completeness limit. The strongest evidence in that sense can
be found in the LFs for the outer Galaxy and for the subcentral
sources, where the $L_\mathrm{min}$ completeness limits are
lowest. The shape for the LF we report is quite different from a power
law functional form (e.g. as used in CT96).

\section{Synthetic fits to the IRAS/CS luminosity function}
\label{sec:model}

One immediate consequence of the shape of the IRAS/CS luminosity
function is that these sources are better understood as clusters of
stars rather than in the framework of one dominant star per source. A
single exciting star would result in a power law LF: an IMF index
$\alpha$ (see below), and mass-luminosity relation $L \propto
M^\gamma$, give a probability distribution in luminosity $p(L) \propto
L^{-\frac{\alpha+\gamma}{\gamma}}$.

We used a simple model to examine whether the variations in the LFs of
IRAS/CS sources from inside to outside the solar circle can be
traced to the underlying young stellar population.  We proceed to
describe a Monte Carlo analysis for the ensemble of massive star
forming regions in the galactic disk.  The luminosity of a MSFR is the
sum of the luminosities of each star, given by the mass-luminosity
relationship. We used a polynomial fit to the mass-luminosity relation
of the tracks presented in Schaller et al. (1992) for Z=0.02, at the
first time step they list,
\begin{eqnarray}\label{eq:M-L}
\log(L/L_{\odot}) = -0.127+4.656 \, \log(M/M_{\odot})\\ \nonumber
-0.764\, \log^2(M/M_{\odot}) .
\end{eqnarray}
The metallicity dependence of the mass-luminosity relation was
neglected, as the Z=0.001 tracks in Schaller et al. (1992) have
luminosities within 10\% of Eq. \ref{eq:M-L} for
$M\,>\,7~M_{\odot}$.

The synthetic population of MSFRs was generated in the following
way. The number of stars in a given MSFR, $N_{\star}$, is generated
randomly within the range $1 \leq N_{\star} \leq
N^\mathrm{max}_{\star}$, and subject to a power law probability
distribution with exponent $-\beta$, $p(N_{\star})\propto
N_{\star}^{-\beta}$. A discussion of the model sensitivity on
$N^\mathrm{max}_{\star}$ will be found in Sect.
\ref{sec:paramspace}. Given $N_{\star}$, the total luminosity of a
MSFR is calculated by summing the individual luminosities of each
star, using the mass-luminosity relationship. The mass of each star is
generated randomly, within the range $1 \leq M/M_{\odot} \leq 120$ and
satisfying the IMF distribution,
\begin{equation}
p(M) \propto M^{-(1+\alpha)};
\end{equation}
in this notation the Salpeter (1955) IMF corresponds to
$\alpha=1.35$. Thus the luminosity of each MSFR is randomly
generated with $L_{\rm MSFR}=\sum_{i=1}^{N_{\star}}L(M_{i})$, and an
ensemble of 5000 MSFRs provides a population large enough to compute
the synthetic luminosity function (that this is one order of magnitude
larger than the number considered in the observed LF does not affect
the results, only helps to reduce the random fluctuations).

An important simplification in this approach is that the ensemble of
MSFRs is assumed to be homogeneous in age (see Sect.
\ref{sec:lifetimes}).  Furthermore, it should be mentioned before
discussing the results of the model that the mass-luminosity relation
remains mainly theoretical for massive stars. Although Burkholder et
al. (1997) give observational evidence that support the massive star
models up to $25\,M_{\odot}$, the upper mass limit we used is
$120\,M_{\odot}$, where to our knowledge no direct observational
information exists to back the theoretical models.


\section{Search in parameter space and the role of $N\mathrm{^{max}_{\star}}$}
\label{sec:paramspace}

\begin{figure}
 \resizebox{8cm}{!}{\epsfig{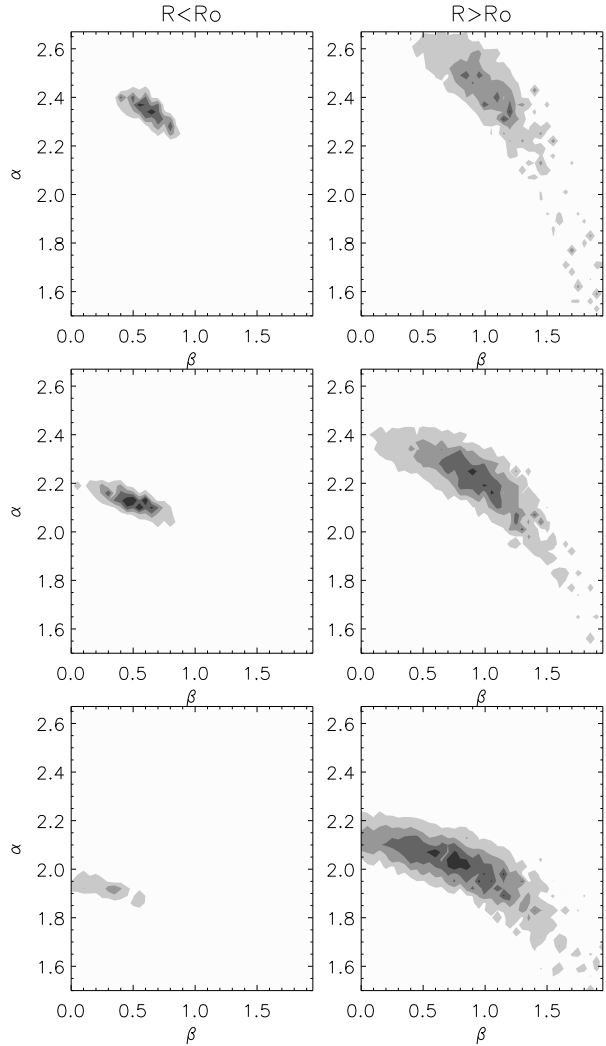}} \caption{Search in
 ($\alpha, \beta$) parameter space for the best fit model, with three
 guesses for $N\mathrm{^{max}_{\star}}$.  Plotted is the $\chi^{2}$
 cumulative probability for the goodness of fit (with 2 free
 parameters), with contours at 10\%, 30\%, 50\% and 70\%.  The upper
 two plots were computed with $N\mathrm{^{max}_{\star}}=2000$, the
 middle two with $N\mathrm{^{max}_{\star}} = 1000$, and the bottom two
 with $N\mathrm{^{max}_{\star}}=500$}\label{fig:paramspace}
\end{figure}

A first broad search for the parameters $\alpha$ and $\beta$ requires
specifying $N\mathrm{^{max}_{\star}}$.  It could be thought firsthand
that any value of $N\mathrm{^{max}_{\star}}$ high enough to simulate
infinity would do, but we tried $N\mathrm{^{max}_{\star}}=10000$ and
did not obtain any fit, over the range $1.5 < \alpha < 2.7$ , $ 0 <
\beta < 2$. We distinguished three cases, fixing the maximum number of
stars per MSFR, $N\mathrm{^{max}_{\star}}$, to 500, 1000 and 2000.
Figure \ref{fig:paramspace} shows the $\chi^{2}$ cumulative
probability for the goodness of fit as a function of parameter
space. In the observed IRAS/CS LF, we have five bins with non-zero
counts above the completeness limit of 10$^{4}\,L_{\odot}$.  The
number of degrees of freedom is thus 3, which corresponds to the five
bins in comparison less two free parameters ($\alpha$ and
$\beta$). The fit to LF$^\mathrm{out}$ is a lot more noisy, a
consequence of the reduced number of IRAS/CS sources used to compute
the LF.

It is apparent that for any value of $N\mathrm{^{max}_{\star}}$,
$\alpha$ is similar inside and outside the solar circle, while the
acceptable values for $\beta$ are markedly different. It can also be
noticed from Fig.~\ref{fig:paramspace} that the best fit $\beta$ are
rather independent of $N\mathrm{^{max}_{\star}}$, in contrast with the
behaviour of the best fit IMF index $\alpha$. For
$N\mathrm{^{max}_{\star}}=1000$, the IMF index in the inner Galaxy, at
50\% confidence level for the goodness of fit (with two free
parameters), corresponds to $2.05 \leq \alpha \leq 2.15$. This is very
close to the result for the outer Galaxy, $2.1 \leq \alpha \leq 2.3$.
The case where $N\mathrm{^{max}_{\star}}=500$ seems to give better
results for $R>R_\circ$, and constrains $1.9<\alpha<2.15$ at 50\%
confidence. This range of values, $\Delta \alpha=0.1$ for $R<R_\circ$
and $\Delta \alpha=0.15$ for $R>R_\circ$, will be used as an
indication of the uncertainty level in the best fit $\alpha$.

As the distribution of $N_{\star}$ seems to be a bounded power law,
there exists a maximum for the number of stars born in a MSFR within a
finite mass range.  It should be kept in mind that the total number of
stars could in fact depend on the stellar masses and the star
formation history of a MSFR; we expect the parameter $N_{\star}$ to
synthesise more complex processes. Observational constraints to fix
$N\mathrm{^{max}_{\star}}$ are difficult to find, because stellar
censuses are available only for much larger regions like OB
associations, open clusters, or longer lived H\,{\sc ii} regions.

\section{The IMF index and  possible large scale variations}
\label{sec:results}

\begin{figure}
 \resizebox{8cm}{!}{\epsfig{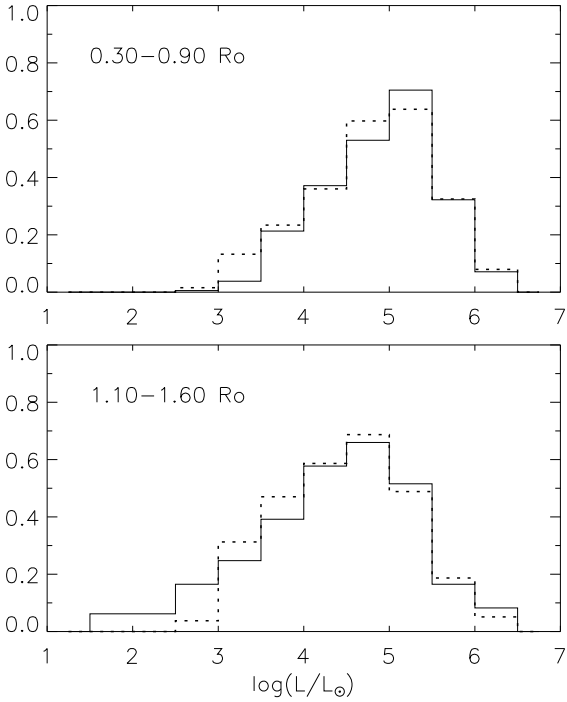}} \caption{Best fit
 models (dotted lines) to the inner and outer Galaxy IRAS/CS LFs
 (solid lines), above the completeness limit $\log(L/L_\odot)=4$. In
 ordinates is the LF normalised over $\log(L/L_\odot)\geq4$, as a
 function of $\log(L/L_{\odot})$}\label{fig:fitLF}
\end{figure}

After finding reasonable values for $N\mathrm{^{max}_{\star}}$, we
extend the search to include all three parameters, and obtain
\begin{eqnarray}
(\alpha, \beta, N\mathrm{^{max}_{\star}})= & (1.988,0.49,646), & {\rm for}~
R<R_{\circ},~ {\rm and}  \\  \nonumber
(\alpha, \beta, N\mathrm{^{max}_{\star}})= & (1.991,0.73,450), & {\rm for} ~ R>R_{\circ}. 
\end{eqnarray}
Thus the best fitting sets of parameters have $\alpha = 2.0$. Values
for the significance of the fits are $\sim60\%$, with two degrees of
freedom (5 bins in comparison less 3 free parameters). Figure
\ref{fig:fitLF} shows the models that best fit the LFs. Due to the
Monte Carlo approach, secondary $\chi^2$ minima are found about the
true minimum. Multiple minimization runs with various initial guesses
fluctuate about the above values: from $(1.98,0.57,708)$ to
$(2.03,0.40,748)$ for $R<R_{\circ}$.

Keeping $N\mathrm{^{max}_{\star}}$ fixed at the best fit value, with
the $\Delta \alpha$ uncertainties quoted in the previous section
(i.e. above a 50\% confidence level in a 2-D slice), our best values
for $\alpha$ are $1.95<\alpha<2.05$ in the inner Galaxy, and
$1.9<\alpha<2.15$ in the outer Galaxy.

As an indication of the uncertainty level related to the use of
Eq.~\ref{eq:Firas}, increasing the bolometric fluxes to $2\, F_{IRAS}$
results in $(\alpha, \beta, N\mathrm{^{max}_{\star}}) =
(1.72,0.29,457)$ for $R<R_{\circ}$. Only very low significance fits
were found using either $5\, F_{IRAS}$, or $0.5\, F_{IRAS}$. A factor
of 2 in the mass-luminosity relation $L(M)$ gives $(\alpha, \beta,
N\mathrm{^{max}_{\star}}) = (2.005,0.48,700)$, i.e. no significant
change. A steep IMF index seems to be a robust result of our
analysis.

The IMF index $\alpha$ for the inner and outer Galaxy seems to be the
same, although $\beta$ changes significantly. In other words, although
the average mass of newly formed stars is the same inside and outside
the solar circle, the average number of stars per star forming region
and in a finite mass range is lower in the outer Galaxy.  The best
fits of the parameter $\beta$ imply that the expectation value for the
number of stars per MSFR with $1 \leq M/M_{\odot} \leq 120$, $<N>$,
decreases from 225 for $R < R_{\circ}$ to 120 for $R > R_{\circ}$. The
decrease in $<N>$ is not an effect related to the area covered by the
100$\,\mu$m IRAS beam: due to the lower FIR background
towards the outer Galaxy, the IRAS point sources are detected to
greater distances than in the inner Galaxy (the average distance to
IRAS/CS sources is 6.92\,kpc for the outer Galaxy, and 5.32\,kpc
in terms of the effective distances for the inner Galaxy).



Is there a gradient in the IMF index with galactocentric radius?  How
does our value for the IMF index compare with other estimates? The
gradient proposed by Garmany et al. (1982) has been re-interpreted as
a result of contamination from field O stars (Massey et al. 1995a),
which seem to derive from a very steep IMF. The values Massey et
al. (1995b) and Massey (1998) quote for $\alpha$ in the Milky Way and
the Magellanic Clouds OB associations are $1.1\pm0.1$ and $1.3\pm0.1$,
which argues against a metallicity dependence of $\alpha$, at least
over 10\,-120\,$M_{\odot}$. But constraining the analysis to nearby OB
associations gives steeper values, Claudius \& Grosb{\o}l (1980) give
$\alpha=1.9$ over 2.2\,-10\,$M_{\odot}$, and Brown (1998) gives
$\alpha$=1.9 for the Upper Scorpius subgroup of Sco OB2, over
3\,-16\,$M_{\odot}$.  Brown (1998) used Hipparcos data to ascertain
membership (see de\,Zeeuw et al. 1999), which is crucial in the coeval
approximation; the very large OB associations in Massey et al.
probably have a complex star formation history. Although Massey et
al. establish a good case for a universal massive star IMF, the
exact value of the IMF index is still an open issue.  Our contribution
to this debate is that there certainly are differences in the physical
processes governing star formation in the outer spiral arms and the
molecular ring. In our simple approach, and understanding that we
restricted our study to regions of massive star formation embedded in
dense molecular cores and with at least one UCH\,{\sc ii} region, it
seems that although the IMF index is constant, the average number of
stars born per region is lower outside the solar circle. We favour a
rather steep IMF index, close to the values from Brown (1998).



\section{Lifetime of UCH\,{\sc ii} regions}\label{sec:lifetimes}

UCH\,{\sc ii} regions are fairly short lived in comparison with giant
H\,{\sc ii} regions. Wood \& Churchwell (1989a) estimated the
lifetimes of UCH\,{\sc ii} regions by comparing the number of O stars
in the solar neighbourhood (Conti et al. 1983) and the number of IRAS
point sources that match the IRAS colours of UCH\,{\sc ii}
regions. They conclude that 10\% to 20\% of an O star's main sequence
lifetime is spent embedded in a molecular cloud, in the UCH\,{\sc ii}
phase. This fraction has been reevaluated to only 0.5\% by CT96, and
our estimate is $\leq 2\%$ (see below). Under the assumption that the
luminosity of a MSFR is dominated by the most massive star, CT96 show
that if there is a variation of UCH\,{\sc ii} lifetime with stellar
mass, it is not as significant as the uncertainties in the IMF. The
crossing time at 10~km\,s$^{-1}$ is 10$^{4}$\,yrs for a typical
UCH\,{\sc ii} region (i.e. G34.3+0.2, which at a distance of 3.7\,kpc
is about $5\,10^{17}\,$cm). Thus the margin between the dynamical
timescale and the time O stars spend in the embedded phase of $<
10^{5}$\,yrs is narrow, and can be affected by many parameters other
than the mass of the most massive star, such as clumpiness of the
molecular clouds and relative motions between the exciting star and
the molecular clumps. It is unlikely that the lifetimes of UCH\,{\sc
ii} regions depend strongly on their stellar contents.

From our synthetic population of MSFRs, we find that the average
number of O stars per MSFR, with luminosities in excess of
$L(20\,M_{\odot}$), is 0.55 for $R < R_{\circ}$ and 0.29 for $R >
R_{\circ}$ - we will consider 0.5 O stars per IRAS/CS
source. Within 2.5 kpc of the Sun, Conti et al. (1983) report 436 O
stars with $M > 20\,M_{\odot}$. We have 15 IRAS/CS sources above
the luminosity limit $L(20\,M_{\odot}$) (counting all sources, outside
the cuts in longitude described in Sect. \ref{sec:obs} but including
the sources with $|V_\mathrm{lsr}| \leq 10~{\rm km\,s}^{-1}$).  The
average fraction of lifetime spent in the embedded phase would thus be
$0.5\times15/(436+0.5\times15)=0.017$.  Alternatively, counting the
IRAS/CS sources with galactocentric radii within 0.9 to
$1.1\,R_{\circ}$, we have 40 sources more luminous than
$L(20\,M_{\odot}$), distributed over an area of 80.5\,kpc$^2$ (BCMN).
This would give 9.8 IRAS/CS sources within 2.5\,kpc of the
Sun. The average fraction of lifetime O stars spend in the embedded
phase would thus be $0.5\times9.8/(436+0.5\times9.8)=0.011$. This
latter approach has the advantage of not suffering as much from the
peculiar velocities of local sources.  But some O stars in the
transition between the embedded UCH\,{\sc ii} phase and the field
stars could have been missed by Conti et al. (1983). Therefore we
estimate that O stars spend $\leq 2\%$ of their lifetime in the
embedded phase.


\section{Conclusions}\label{sec:conclusion}

We have constructed the FIR luminosity function for IRAS point
sources with colours of UCH\,{\sc ii} regions and CS(2--1) detections,
for different sectors in the galactic disk. The LFs are reliable above
a luminosity of $10^4~L_{\odot}$, and extend to $\sim10^6~L_{\odot}$
in the high luminosity end, with a peak at $\sim10^5~L_{\odot}$.  We
analysed the trends in the LF of IRAS/CS sources in terms of an
ensemble of massive star forming regions, and two free parameters
suffice to provide a remarkably good fit to the shape of the LF. The
fits required a maximum for the number of stars born in a given MSFR
with $1 \leq M /M_{\odot} \leq 120$. The best results were obtained
setting $N\mathrm{^{max}_{\star}}=646$ for $R < R_{\circ}$, and
$N\mathrm{^{max}_{\star}}=450$ for $R > R_{\circ}$.  A few conclusions
can be summarised:
 
\begin{enumerate}

\item 
The LFs inside and outside the solar circle are different at 96\%
confidence level. The LF within the solar circle, built with 413
sources, peaks at $2\,10^5L_{\odot}$, while the LF outside the solar
circle, built with 142 sources, peaks at $5\,10^4\,L_{\odot}$.

 
\item 
The IMF index we obtain is $\alpha=2$ and constant with galactocentric
radius. At 50\% cumulative probability for the goodness of fit, and
keeping $N\mathrm{^{max}_{\star}}$ fixed at the best fit value,
$1.95<\alpha<2.05$ in the inner Galaxy, and $1.9<\alpha<2.15$ in the
outer Galaxy.

\item
A power law distribution for the number of stars per MSFR has an
exponent $-\beta \sim -0.49$ in the inner Galaxy. But in the outer
Galaxy the best fit model corresponds to $-\beta \sim -0.73$.  Thus,
the expectation value for the number of stars per MSFR with $1 \leq M
/M_{\odot} \leq 120$ decreases from 225 for $R < R_{\circ}$ to 120 for
$R > R_{\circ}$.

\end{enumerate} 

The results of our analysis show that the observed luminosity
functions for UCH\,{\sc ii} regions can be traced to the underlying young
population. The differences within and outside the solar circle
reflect a decrease in the average number of stars per massive star
forming region towards the outer Galaxy, rather than a steeper IMF.


\begin{acknowledgements}
This article benefited from the suggestions and encouraging comments
of the referee, James Lequeux. We are also grateful to Rodrigo Soto
and Mark Seaborne for helpful discussions.  The staff at the SEST and
OSO telescopes kindly assisted us within the course of the
observations.  The Swedish-ESO Submillimetre Telescope is operated
jointly by ESO and the Swedish National Facility for Radioastronomy,
Onsala Space Observatory, at Chalmers University of Technology. The
Onsala 20m telescope is operated by the Swedish National Facility for
Radioastronomy, Onsala Space Observatory, at Chalmers University of
Technology. S.C. acknowledges support from Fundaci\'{o}n Andes and
PPARC through a Gemini studentship.  L.B., S.C., and J.M. acknowledge
support from FONDECYT-Chile grant 8970017, and from a C\'atedra
Presidencial en Ciencias 1997.
\end{acknowledgements}

\end{document}